\newcommand{\version}{May 30, 2013}
\documentclass[a4paper,twoside,11pt,pdftex]{article}
\usepackage[bindingoffset=0.4cm,textheight=22.5cm,hdivide={2.7cm,*,2.7cm}, vdivide={*,22cm,*}]{geometry}
\usepackage{amsmath,amsfonts,amssymb,slashed}
\IfFileExists{dsfont.sty}
	{\usepackage{dsfont}
         \let\mathbb=\mathds
         \newcommand{\id}{\mathds{1}}}
	{\typeout{Package dsfont.sty was not found, using alternative macros.}
         \let\mathds=\mathbb
         \newcommand{\id}{\mbox{1 \kern-.59em {\rm l}}}}
\DeclareMathOperator{\tr}{Tr} 
\DeclareMathOperator{\Tr}{Tr}
\usepackage[pdftex]{graphicx}
\usepackage[pdftex,bookmarksnumbered=true,breaklinks=true]{hyperref}

\usepackage[numbers,square,comma,sort&compress]{natbib}

\newcommand{\uim}{UV/IR mixing}
\newcommand{\nc}{non-com\-mu\-ta\-tive}

\newcommand{\nn}{\nonumber}
\newcommand{\eqnref}[1]{Eqn. (\ref{#1})}		
\newcommand{\figref}[1]{Fig.~\ref{#1}}			
\newcommand{\secref}[1]{Section~\ref{#1}}		
\newcommand{\appref}[1]{Appendix~\ref{#1}}		
\newcommand{\co}[2]{\left[#1,#2\right]}					
\newcommand{\aco}[2]{\left\{#1,#2\right\}}				
\newcommand{\starco}[2]{\left[ #1\stackrel{\star}{,}#2\right] }		
\newcommand{\var}[2]{\frac{\d #1}{\d #2}}				
\newcommand{\vvar}[3]{\frac{\d^2 #1}{\d #2\d #3}}			
\newcommand{\vvvar}[4]{\frac{\d^3 #1}{\d #2\d #3\d #4}}
\newcommand{\vvvvar}[5]{\frac{\d^4 #1}{\d #2\d #3\d #4\d #5}}
\newcommand{\pa}{\partial}						
\newcommand{\diff}[2]{\frac{\pa #1}{\pa #2}}				
\newcommand{\ddiff}[3]{\frac{\partial^2 #1}{\partial #2 \partial #3}}   
\newcommand{\ri}{\mathrm{i}}						
\newcommand{\re}{\mathrm{e}}						
\renewcommand{\k}{\tilde{k}}						
\newcommand{\p}{\tilde{p}}						
\newcommand{\bc}{\bar{c}}						
\newcommand{\Gam}{\Gamma^{(0)}}						
\newcommand{\Act}{S}
\renewcommand{\a}{\alpha}
\renewcommand{\b}{\beta}
\newcommand{\g}{\gamma}
\renewcommand{\d}{\delta}
\newcommand{\e}{\epsilon}

\renewcommand{\th}{\theta}
\newcommand{\mth}{\theta} 
\newcommand{\sth}{\varepsilon} 

\newcommand{\m}{\mu}
\newcommand{\n}{\nu}

\renewcommand{\r}{\rho}
\newcommand{\s}{\sigma}
\renewcommand{\t}{\tau}

\newcommand{\w}{\omega}

\renewcommand{\L}{\Lambda}

\newcommand{\W}{\Omega}

\newcommand{\inv}[1]{\frac1{#1}}				
\newcommand{\tinv}[1]{\tfrac{1}{#1}}
\newcommand{\intx}{\int\!\! {\rm d}^4x}						
\newcommand{\wsq}{\widetilde{\square}}
\newcommand{\ig}{\mathrm{i}g}

\newcommand{\bpsi}{\bar{\psi}}
\newcommand{\bB}{\bar{B}}
\newcommand{\bQ}{\bar{Q}}
\newcommand{\bJ}{\bar{J}}

\newcommand{\mF}{\mathcal{F}}
\newcommand{\R}{\mathds{R}}

\title{\begin{flushright}
        {\small LA-UR-13-20923}
       \end{flushright}\vspace{3em}
Slavnov-Taylor identities, {\nc} gauge theories and infrared divergences}
\author{Daniel N. Blaschke\footnote{\ttfamily{dblaschke@lanl.gov}}~, Harald Grosse\footnote{\ttfamily{harald.grosse@univie.ac.at}}~ and Jean-Christophe Wallet\footnote{\ttfamily{jean-christophe.wallet@th.u-psud.fr}}}

\date{\version}
\graphicspath{{./}{./figures/}}
\begin{document}
\maketitle
\thispagestyle{empty}
\begin{center}
\renewcommand{\thefootnote}{\fnsymbol{footnote}}
\vspace{-0.3cm}
\footnotemark[1]\textit{Theory Division, Los Alamos National Laboratory\\
Los Alamos, NM, 87545, USA}\\[0.3cm]
\footnotemark[2]\textit{University of Vienna, Faculty of Physics\\
Boltzmanngasse 5, A-1090 Vienna, Austria}\\[0.3cm]
\footnotemark[3]\textit{Laboratoire de Physique Th\'eorique, B\^at.\ 210\\
CNRS and Universit\'e Paris-Sud 11,  91405 Orsay Cedex, France}\\[0.3cm]
\end{center}%
\begin{abstract}
In this work we clarify some properties of the one-loop IR divergences in non-Abelian gauge field theories on {\nc} 4-dimensional Moyal space. 
Additionally, we derive the tree-level Slavnov-Taylor identities relating the two, three and four point functions, and verify their consistency with the divergent one-loop level results.
We also discuss the special case of two dimensions.
\end{abstract}

\newpage
\tableofcontents

\section{Introduction}

Quantum field theories formulated on {\nc} spaces are motivated by the fact that the classical concept of space and time must break down at Planck scale distances.
The simplest example of a {\nc} space is achieved by deforming Euclidean space by assuming that its coordinates fulfill a Heisenberg algebra $[\hat x^\m,\hat x^\n]$, thereby promoting them to operators on a Hilbert space.
A simple realization of this so-called Groenewold-Moyal deformed (or $\th$-deformed) space~\cite{Groenewold:1946,Moyal:1949} is formulated in terms of ordinary functions by means of a deformed star-product
\begin{align}
(f\star g) (x) = \re^{\ri\frac{\sth}2 \mth_{\mu\nu} \partial^x_\mu \partial^y_\nu } f(x) g(y) \Big|_{y\to x}
\,.
\end{align}
The real parameter $\sth$ has mass dimension $-2$ rendering the constant antisymmetric deformation matrix $\mth_{\mu\nu}$ dimensionless. 
One can easily check that
\begin{align}
\starco{x_\mu}{x_\nu} = \ri \sth \mth_{\mu\nu}
\,.
\end{align}
This commutation relation is invariant under translations of the space-time coordinates and under the so-called reduced Lorentz transformations (or reduced orthogonal transformations in the Euclidean setting) see e.g.~\cite{Grosse:2011es}.
For a discussion on general properties of star-products oriented to physics, see \cite{GraciaBondia:2001ct}.
Reviews on quantum field theories (QFTs) on Groenewold-Moyal deformed spaces may be found in e.g.~\cite{Szabo:2001,Rivasseau:2007a,Wallet:2007em,Blaschke:2010kw}.

In general, such models suffer from new types of divergences arising due to a phenomenon referred to as {\uim}~\cite{Minwalla:1999px,Matusis:2000jf}.
Only some years ago, Grosse and Wulkenhaar were able to resolve this problem in the case of a scalar field theory with quartic coupling by adding a harmonic oscillator-like term to the (Euclidean) action thereby rendering it renormalizable to all orders of perturbation theory~\cite{Grosse:2003, Grosse:2004b}. The Grosse-Wulkenhaar model has a vanishing $\beta$-function to all orders \cite{Grosse:2004a,Disertori:2006uy,Rivasseau:2006b} when the action is invariant under the Langmann-Szabo duality \cite{Langmann:2002}. It is very likely to be non-perturbatively solvable, as shown in \cite{Grosse:2012uv}. Besides, this model as well as its gauge theory counterpart (at the classical level) is linked with a peculiar type of spectral triple \cite{Grosse:2007jy} whose relationship to the Moyal geometries has been analyzed in \cite{Wallet:2011aa,Cagnache:2009xj,Cagnache:2009ik}. The harmonic term admits a geometric interpretation in terms of {\nc} scalar curvature~\cite{Buric:2009ss,Buric:2010,deGoursac:2010zb}. 
A variation of the Grosse-Wulkenhaar model tailored to degenerate Moyal space was explored in~\cite{Grosse:2008df}.

Later, an alternative approach was put forward in \cite{Gurau:2009} by replacing the oscillator term with a translation invariant alternative of type $\phi(-p)\inv{p^2}\phi(p)$.
Also in this case, the authors were able to prove renormalizability of this ``$1/{p^2}$-model'' to all orders by means of Multiscale Analysis. The restoration of rotational invariance while preserving renormalizability of the Grosse-Wulkenhaar model has been discussed in \cite{deGoursac:2009un}.

Inspired by these successes, similar approaches were examined for $U_\star(1)$ gauge theories\footnote{By the subscript $\star$ we emphasize that the non-commutativity of the space coordinates alters the gauge group.} in Euclidean space:
A gauge model induced by Heatkernel methods of the Grosse-Wulkenhaar model coupled to  a gauge field was first employed in~\cite{Grosse:2007,Wallet:2007c}.
However, its non-trivial vacuum structure poses thus far unresolved problems~\cite{Grosse:2007jy,deGoursac:2007uv,Wallet:2008a}.
An alternative route of implementing an oscillator term in a {\nc} gauge theory was tried in~\cite{Blaschke:2007b}, but was found to generate the same induced model at one-loop order~\cite{Blaschke:2009aw}.
Various approaches to implementing a damping mechanism similar to the scalar $1/p^2$-model were also discussed --- see~\cite{Blaschke:2008a,Vilar:2009,Blaschke:2009e,Blaschke:2010ck}.
Classical structures stemming from non-commutative differential geometry that may underly these gauge models have been explored in \cite{Marmo:2004re,Wallet:2008bq,Wallet:2008b}.
In \secref{sec:comments} we will comment on the approach of Ref.~\cite{Blaschke:2010ck}.

Finally, methods and problems arising when one attempts to prove renormalizability of a gauge model on {\nc} spaces can be found in~\cite{Blaschke:2009c,Blaschke:2012ex} and references therein.

The main purpose of this work, however, will be to clarify some properties of the one-loop IR divergences of non-Abelian gauge field theories on Moyal-deformed spaces.
We therefore start by discussing explicit one-loop calculations on {\nc} $\R^4_\th$ after ``setting the stage'' by introducing the model and its properties, such as Slavnov-Taylor identities, in \secref{sec:modelproperties} and \ref{sec:oneloop}.
We then make some comments concerning its renormalizability in \secref{sec:comments} and finally discuss the special case of two dimensions in \secref{sec:2d}.

\paragraph{Notation.}
Throughout the remainder of this paper, the following notation will be used: 
Following Ref.~\cite{Armoni:2000xr} we denote $U_\star(N)$ indices with capital letters $A,B,C,\ldots$ and $SU_\star(N)$ indices with $a,b,c,\ldots$. Finally, the index $0$ is used for fields which are $U_\star(1)$, and whenever an index is omitted, the according field including the $U(N)$ gauge group generator $T^A$ is meant. 
Furthermore, we implicitly assume all products to be deformed (i.e. star pro\-ducts). 
Finally, we define the following contractions with $\mth_{\mu\nu}$:
\begin{align}
\tilde v_\mu := \mth_{\mu\nu} v_\nu \,, \qquad \tilde w := \mth_{\mu\nu} w_{\mu\nu}\,.
\end{align}

\paragraph{$\mathbf{U_\star(N)}$ gauge fields.}
The covariant derivative $D_\m$ and the field strength $F_{\m\n}$ are defined as
\begin{align}
D_\m\bullet&=\pa_\m\bullet-\ig\co{A_\m}{\bullet}\,, & A_\m&=A_\m^AT^A\,, \nonumber\\
F_{\m\n}&=\pa_\m A_\n-\pa_\n A_\m-\ig\co{A_\m}{A_\n}\,,
\end{align}
where $T^A$ are the generators of the $U(N)$ gauge group. They are normalized as $\Tr(T^AT^B)=\inv{2}\d^{AB}$, and $T^0=\inv{\sqrt{2N}}\id_N$ (cf.~\cite{Armoni:2000xr}). 
Due to the star product, the field strength tensor $F_{\m\n}$ exhibits additional couplings between the $U_\star(1)$ and the $SU_\star(N)$ sector, i.e. we have
\begin{align}
F_{\m\n}&=\left(\pa_\m A^0_\n-\pa_\n A^0_\m-\frac{\ig}{2}d^{AB0}\co{A^A_\m}{A^B_\n}\right)T^0\nonumber\\
&\quad +\left(\pa_\m A^c_\n-\pa_\n A^c_\m+\frac{g}{2}f^{abc}\aco{A^a_\m}{A^b_\n}-\frac{\ig}{2}d^{ABc}\co{A^A_\m}{A^B_\n}\right)T^c\nonumber\\
&\equiv F^0_{\m\n}T^0+F^c_{\m\n}T^c \,,
\end{align}
where $f^{abc}$ and $d^{ABC}$ are (anti)symmetric structure constants of the gauge group. 
The terms proportional to $d^{AB0}=\sqrt{\frac{2}{N}}\d^{AB}$ contain both types of fields, i.e. $U_\star(1)$ and $SU_\star(N)$, and hence giving rise to the additional couplings. 
In the commutative limit, the star commutators would vanish and the two sectors would decouple once more.

Similarly, one has for the covariant derivative of e.g. a ghost field $c$:
\begin{align}
D_\m c&=\left(\pa_\m c^0-\frac{\ig}{2}d^{AB0}\co{A^A_\m}{c^B}\right)T^0 +\left(\pa_\m c^c+\frac{g}{2}f^{abc}\aco{A^a_\m}{c^b}-\frac{\ig}{2}d^{ABc}\co{A^A_\m}{c^B}\right)T^c
\,.
\end{align}

\section{Non-commutative gauge field action and its symmetries}
\label{sec:modelproperties}
We consider the {\nc} $U(N)$ gauge field action on Euclidean $\R^4_\th$ with a covariant gauge fixing
\begin{align}
S&=\Tr\intx\left(\inv{4}F_{\m\n} F^{\m\n}+s\left(\bc \pa^\m A_\m+\tfrac{\a}{2}\bc b\right)\right) \nn\\
&=\Tr\intx\left(\inv{4}F_{\m\n} F^{\m\n}+b \pa^\m A_\m +\tfrac{\a}{2}b b-\bc\pa^\m D_\m c\right)
\,, \label{eq:naive-gauge-action}
\end{align}
which is invariant under the BRST transformations
\begin{align}
s A_\mu & =  D_\mu c\,, &
sc & = \ig c c \,, \nn\\
s\bar c & = b\,, & s b &= 0 \,, \nn\\
s^2 \phi & = 0\,, \quad \forall \phi 
\,. \label{eq:BRST-naive-NC-gauge}
\end{align}
All products are considered to be star products.
Upon introducing external sources for non-linear BRST-transformations
\begin{align}
 \Gam&=S+S_{\text{ext}}\,, &
 S_{\text{ext}}&=\Tr\intx\left(\r^\m sA_\m+\s sc\right)
 \,,
\end{align}
one derives the identity
\begin{align}
 \mathcal{S}(\Gam)&=\intx\left(\var{\Gam}{\r^{\m,A}(x)}\var{\Gam}{A^A_\m(x)}+\var{\Gam}{\s^A(x)}\var{\Gam}{c^A(x)}+b^A(x)\var{\Gam}{\bc^A(x)}\right)=0
 \,, \label{eq:id-1}
\end{align}
capturing the (BRST) symmetry content of the model at tree level.
The according linearized symmetry operator then reads
\begin{align}
 \mathcal{S}_{\Gam}&=\intx\left(\var{\Gam}{\r^{\m,A}(x)}\var{\ }{A^A_\m(x)}+\var{\Gam}{A^A_\m(x)}\var{\ }{\r^{\m,A}(x)}+\var{\Gam}{\s^A(x)}\var{\ }{c^A(x)} \right.\nn\\
 &\qquad\qquad \left. +\var{\Gam}{c^A(x)}\var{\ }{\s^A(x)}+b^A(x)\var{\ }{\bc^A(x)}\right)
 \,, \label{eq:lin-op}
\end{align}
Noting that
\begin{align}
 \var{\ }{A^A_\n(y)}\mathcal{S}(\Gam)&=\mathcal{S}_{\Gam}\var{\Gam}{A^A_\n(y)}=0
 \,,
\end{align}
we can vary this last relation with respect to $c(z)$ and set all fields to zero afterwards resulting in
\begin{align}
 \pa^z_\m\vvar{\Gam}{A^A_\n(y)}{A^B_\m(z)}\Big|_{\Phi=0}&=0
 \,. \label{eq:id-2}
\end{align}
This implies transversality of the two point function.

Similarly, in varying an additional time with respect to $A$ one derives an identity relating the 3-point function and the 2-point function, i.e.
\begin{align}
 \vvvar{\mathcal{S}(\Gam)}{c^C\!(z)}{A^A_\s(x)}{A^B_\n(y)}&=0
 \,,
\end{align}
leads (for vanishing fields) to the tree-level identity
\begin{align}
 \pa^z_\m\vvvar{\Gam}{A^A_\s(x)}{A^B_\n(y)}{A^C_\m(z)}&=\ig d^{DAC}\!\co{\vvar{\Gam}{A^D_\s(x)}{A^B_\n(y)}}{\d(y-z)}\nn\\
 &\quad +\ig f^{DAC}\!\aco{\vvar{\Gam}{A^D_\s(x)}{A^B_\n(y)}}{\d(y-z)}+(\s,A,x)\leftrightarrow(\n,B,y)
 \,, \label{eq:id-3}
\end{align}
where the star-product is with respect to the variable that appears in both the 2-point function and the delta-function (i.e. $y$ in the first two terms and $x$ in the other terms on the rhs).

Finally, one additional variation with respect to the gauge field yields an identity relating the four to the three point function:
\begin{align}
\pa^w_\m\vvvvar{\Gam}{A^A_\e(z)}{A^B_\s(x)}{A^C_\n(y)}{A^D_\m(w)}&=\ig d^{EBD}\co{\vvvar{\Gam}{A^A_\e(z)}{A^E_\s(x)}{A^C_\n(y)}}{\d(y-w)}\nn\\
&\quad+\ig f^{EBD}\aco{\vvvar{\Gam}{A^A_\e(z)}{A^E_\s(x)}{A^C_\n(y)}}{\d(y-w)}\nn\\
&\quad+(\s,B,x)\leftrightarrow(\n,C,y)+(\e,A,z)\leftrightarrow(\n,C,y)
 \,. \label{eq:id-4}
\end{align}
One of the goals of this paper will be to check these identities for the IR divergences appearing at one-loop level.

\section{One-loop computations}
\label{sec:oneloop}
\subsection{Feynman rules}
The gauge field propagator takes the form
\begin{align}
 G^{A^AA^B}_{\m\n}(k)&=\frac{\d^{AB}}{k^2}\left(\d_{\m\n}-(1-\a)\frac{k_\m k_\n}{k^2}\right)
\,, \label{eq:propA}
\end{align}
and the ghost propagator takes the simple form
\begin{align}
G^{\bc c}(k)=-\frac{\d^{AB}}{k^2}
\,. \label{eq:prop_cbc}
\end{align}
Additionally, the model \eqref{eq:naive-gauge-action} features several vertices:
{\allowdisplaybreaks
\begin{subequations}
\begin{align}
\hspace{-3ex}\raisebox{-20pt}{\includegraphics[scale=0.8,trim=0 5pt 0 5pt,clip=true]{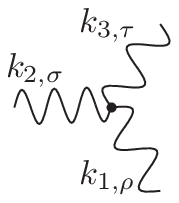}}&\widetilde{V}^{A^AA^BA^0}_{\rho\s\tau}(k_1, k_2, k_3)=2\ig(2\pi)^4\d^4(k_1+k_2+k_3)\mF^{AB0}(k_1,k_2)\times\nonumber\\[-12pt]
&\hspace{3.2cm}\quad\times\left[(k_3-k_2)_\rho \d_{\s\tau}+(k_1-k_3)_\s \d_{\rho\tau}+(k_2-k_1)_\tau \d_{\rho\s}\right],\nonumber\\*
&\widetilde{V}^{A^aA^bA^c}_{\rho\s\tau}(k_1, k_2, k_3)=2\ig(2\pi)^4\d^4(k_1+k_2+k_3)\mF^{abc}(k_1,k_2)\times\nonumber\\
&\hspace{3.2cm}\quad\times\left[(k_3-k_2)_\rho \d_{\s\tau}+(k_1-k_3)_\s \d_{\rho\tau}+(k_2-k_1)_\tau \d_{\rho\s}\right],\label{eq:vert_3a}\\
\hspace{-3ex}\raisebox{-23pt}{\includegraphics[scale=0.8,trim=0 2pt 0 5pt,clip=true]{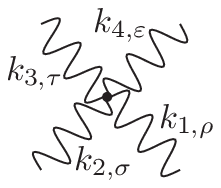}}\hspace{-9pt}&\widetilde{V}^{4A}_{\rho\s\tau\e}(k_1, k_2, k_3, k_4)=-4g^2(2\pi)^4\d^4(k_1+k_2+k_3+k_4)\times\nonumber\\*[-16pt]
&\hspace{4cm}\times\Big[(\d_{\rho\tau}\d_{\s\e}-\d_{\rho\e}\d_{\s\tau})\mF^{ABE}(k_1,k_2)\mF^{CDE}(k_3,k_4)\nonumber\\*
&\hspace{4.5cm}+(\d_{\rho\s}\d_{\tau\e}-\d_{\rho\e}\d_{\s\tau})\mF^{ACE}(k_1,k_3)\mF^{BDE}(k_2,k_4)\nonumber\\*
&\hspace{4.5cm}+(\d_{\rho\s}\d_{\tau\e}-\d_{\rho\tau}\d_{\s\e})\mF^{BCE}(k_2,k_3)\mF^{ADE}(k_1,k_4)\Big],\label{eq:vert_4a}\\
\hspace{-3ex}\raisebox{-23pt}{\includegraphics[scale=0.8,trim=0 2pt 0 5pt,clip=true]{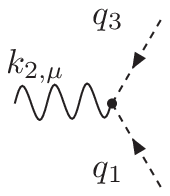}}\hspace{0.1ex}&\widetilde{V}^{\bc^0 A^Ac^B}_\mu(q_1, k_2, q_3)=-2\ig(2\pi)^4\d^4(q_1+k_2+q_3)q_{3\mu}\mF^{AB0}(q_1,q_3),\nonumber\\*[-15pt]
&\widetilde{V}^{\bc^a A^bc^c}_\mu(q_1, k_2, q_3)=-2\ig(2\pi)^4\d^4(q_1+k_2+q_3)q_{3\mu}\mF^{acb}(q_1,q_3) 
\,, \label{eq:vert_cAbc}
\end{align}
\end{subequations}
with
}
\begin{align}
\mF^{abc}(k_i,k_j)&=\left(\tfrac{d^{abc}}{2}\sin\!\left(\tfrac{\sth}{2} k_i\k_j\right)+\tfrac{f^{abc}}{2}\cos\!\left(\tfrac{\sth}{2} k_i\k_j\right)\right)\,,\nonumber\\
\mF^{AB0}(k_i,k_j)&=\tfrac{d^{AB0}}{2}\sin\!\left(\tfrac{\sth}{2} k_i\k_j\right) =\tfrac{\d^{AB}}{\sqrt{2N}}\sin\!\left(\tfrac{\sth}{2} k_i\k_j\right) \,, \nn\\
\mF^{a00}(k_i,k_j)&=0
\,. \label{eq:phase-factors}
\end{align}
Considering the scaling behaviour of all these Feynman rules for large momenta, one easily derives an estimate for the superficial degree of ultraviolet divergences:
\begin{align}\label{eq:powercounting}
 d_\gamma=4-E_A-E_{c\bc}\,,
\end{align}
where $E$ denotes the number of external legs of the various field types in a Feynman graph. Having derived all necessary tools (see also the identities given in Appendix~\ref{app:identities}), the various one-loop corrections may be computed. Their results are presented in the following sections.

\subsection{Vacuum polarization}\label{sec:vac-pol}
\begin{figure}[!ht]
 \centering
 \includegraphics[scale=0.8]{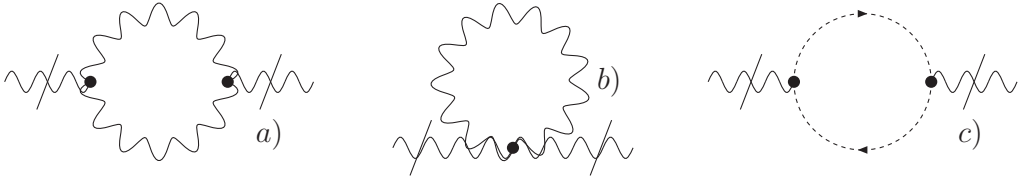}
 \caption{One loop corrections to the gauge boson propagator.} 
 \label{fig:vacpol_all}
\end{figure}

The Feynman rules of the previous section give rise to three graphs contributing to the vacuum polarization which are depicted in Figure~\ref{fig:vacpol_all}. 
However, we have to distinguish between the cases where the external (amputated) legs are $U(1)$ and where they belong to the $SU(N)$ subsector. In the first case we find a quadratic IR divergence of the form
\begin{align}
\Pi^{\text{IR}}\big(A^0_\m(p)A^0_\n(-p)\big)&=\frac{Ng^2}{\pi^2}\frac{\p_\m\p_\n}{\sth^2(\p^2)^2}+\text{finite},
\label{eq:vac-pol-IR-U1}
\end{align}
and a logarithmic UV divergence
\begin{align}\label{eq:vac-pol-UV-U1}
\Pi^{\text{UV}}\big(A^0_\m(p)A^0_\n(-p)\big)&=\frac{(13-3\a)Ng^2}{96\pi^2}\left(p^2\d_{\m\n}-p_\m p_\n\right)\ln|\L^2\sth^2\p^2|+\text{finite}.
\end{align}
For the case of only external $SU(N)$ legs, on the other hand, each of the graphs depicted in \figref{fig:vacpol_all} is strictly planar, meaning their non-planar contributions are zero and hence IR finite (which in fact is consistent with the result of~\cite{Armoni:2000xr}).
In fact, this can be easily seen by considering the according phase factors when the free colour indices $a,b\in SU_\star(N)$.
In that case, one has phase factors of the form
\begin{align}
d^{aCD}d^{bCD}\sin^2\!\left(\tfrac{\sth}{2}k\p\right)+f^{acd}f^{bcd}\cos^2\!\left(\tfrac{\sth}{2}k\p\right)=N\d^{ab}\,,
\label{eq:SUN-phases-vacpol}
\end{align}
since $d^{aCD}d^{bCD}=f^{acd}f^{bcd}=N\d^{ab}$. Clearly, they are phase-independent and hence lead to purely planar contributions. 

The sum of (planar) graphs, however, is logarithmically UV divergent exhibiting exactly the same numerical factor as \eqref{eq:vac-pol-UV-U1}, i.e.
\begin{align}
\Pi^{\text{UV}}\big(A^a_\m(p)A^b_\n(-p)\big)&=\d^{ab}\frac{(13-3\a)Ng^2}{96\pi^2}\left(p^2\d_{\m\n}-p_\m p_\n\right)\ln|\L^2\sth^2\p^2|+\text{finite}.
\label{eq:vacpol-sun}
\end{align}
Furthermore, all three results, \eqref{eq:vac-pol-IR-U1}, \eqref{eq:vac-pol-UV-U1} and \eqref{eq:vacpol-sun}, are transverse with respect to $p^\m$ in accordance with the Slavnov-Taylor identity \eqref{eq:id-2}.

\subsection{\texorpdfstring{$3A$}{3A}-vertex corrections}
\begin{figure}[!ht]
 \centering
 \includegraphics[scale=0.8]{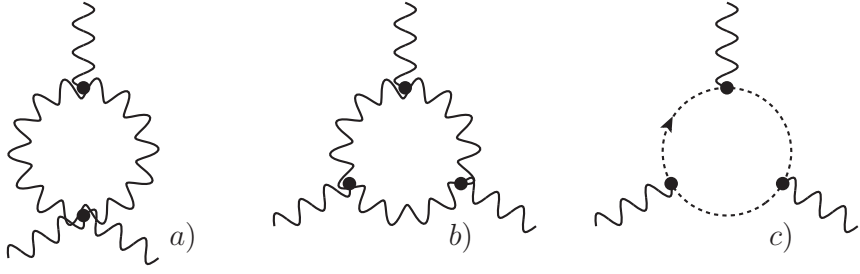}
 \caption{One loop corrections to the 3A-vertex.} 
 \label{fig:1loop_3A_all}
\end{figure}

There are essentially three different types of graphs contributing to the $3A$ vertex corrections at one-loop level. These are depicted in \figref{fig:1loop_3A_all}. 
Useful identities for the structure constants are given in \appref{app:identities}, and through explicit calculations we find the following results:

\subsubsection*{Only external $U(1)$ legs}
When considering only external $U(1)$ legs, the sum of the graphs depicted in \figref{fig:1loop_3A_all} yields a linear IR divergence of the form
\begin{align}\label{eq:3A_correction_IR_U1}
\Gamma^{3A^0,\text{IR}}_{\m\n\r}(p_1,p_2,p_3)&=-\frac{\ri g^3}{\pi^2}\sqrt{\frac{N}{2}}\cos\left(\sth \frac{p_1\p_2}{2}\right)\sum\limits_{i=1,2,3}\frac{\p_{i,\m}\p_{i,\n}\p_{i,\r}}{\sth(\p_i^2)^2}\,,
\end{align}
as well as a logarithmic UV divergence
\begin{align}\label{3A_correction_UV_U1}
\Gamma^{3A^0,\text{UV}}_{\m\n\r}(p_1,p_2,p_3)&=\frac{(17-9\a)}{3(4\pi)^2}\ig^3\sqrt{\frac{N}{2}}\ln(\L)\sin\!\left(\!\sth \frac{p_1\p_2}{2}\right)\Big[(p_1-p_2)_\r\d_{\m\n}+(p_2-p_3)_\m\d_{\n\r}\nonumber\\
&\hspace{4.9cm}+(p_3-p_1)_\n\d_{\m\r}\Big]\nonumber\\
&=-\frac{(17-9\a)\,g^2N}{6(4\pi)^2}\ln (\Lambda) \widetilde{V}^{3A^0,\text{tree}}_{\m\n\r}(p_1,p_2,p_3)
\,. 
\end{align}
Note, that the corresponding would-be logarithmic IR divergence is actually finite for small $\p$ due to the combination\footnote{Since all external legs are in the $U(1)$, the cosines drop out.)} $\sin\!\left(\!\sth \frac{p_1\p_2}{2}\right)\ln(\sth^2\p_i^2)\approx \sth \frac{p_1\p_2}{2}\ln(\sth^2\p_i^2)\to0$, where $i=1,2,3$ and $p_1+p_2+p_3=0$.  
These results are in agreement with the literature~\cite{Matusis:2000jf, Armoni:2000xr,Ruiz:2000}.

\subsubsection*{Only external $SU(N)$ legs}
The sum of these graphs only exhibit a logarithmic UV divergence
\begin{align}\label{3A_correction_UV_SUN}
\Gamma^{A^aA^bA^c,\text{UV}}_{\m\n\r}(p_1,p_2,p_3)&=\frac{(17-9\a)}{3(4\pi)^2}\ig^3\pi^2N\ln(\L)\left(\frac{d^{abc}}{2}\sin\left(\sth \frac{p_1\p_2}{2}\right)+\frac{f^{abc}}{2}\cos\left(\sth \frac{p_1\p_2}{2}\right)\right)\nonumber\\
&\quad\quad \times\Big[(p_1-p_2)_\r\d_{\m\n}+(p_2-p_3)_\m\d_{\n\r}+(p_3-p_1)_\n\d_{\m\r}\Big]\nonumber\\
&=-\frac{(17-9\a)\,g^2N}{6(4\pi)^2}\ln (\Lambda) \widetilde{V}^{A^aA^bA^c,\text{tree}}_{\m\n\r}(p_1,p_2,p_3)
\,.
\end{align}
However, there is no IR divergent part (neither linear nor logarithmic) in this case: Every one of the three contributing one-loop graphs of \figref{fig:1loop_3A_all} (where all internal lines denote the full $U(N)$ propagators) is IR finite. 
This is consistent with the results of~\cite{Armoni:2000xr}.

\subsubsection*{One external $U(1)$ leg and two external $SU(N)$ legs}
Once more, one finds a logarithmic UV divergence
\begin{align}\label{3A_correction_UV_mixed}
\Gamma^{A^aA^bA^0,\text{UV}}_{\m\n\r}(p_1,p_2,p_3)&=\frac{(17-9\a)}{3(4\pi)^2}\ig^3\ln(\L)\sqrt{\frac{N}{2}}\d^{ab}\sin\left(\sth \frac{p_1\p_2}{2}\right)\nonumber\\
&\quad\quad \times\Big[(p_1-p_2)_\r\d_{\m\n}+(p_2-p_3)_\m\d_{\n\r}+(p_3-p_1)_\n\d_{\m\r}\Big]\nonumber\\
&=-\frac{(17-9\a)\,g^2N}{6(4\pi)^2}\ln (\Lambda) \widetilde{V}^{A^aA^bA^0,\text{tree}}_{\m\n\r}(p_1,p_2,p_3)
\,, 
\end{align}
and a linear IR divergence
\begin{align}
\Gamma^{A^aA^bA^0,\text{IR}}_{\m\n\r}\big(p_1,p_2,p_3\big)&=-\frac{\ri g^3}{\pi^2}\sqrt{\frac{N}{2}}\d^{ab}\cos\left(\sth \frac{p_1\p_2}{2}\right)\frac{\p_{3,\m}\p_{3,\n}\p_{3,\r}}{\sth(\p_3^2)^2}
\,, \label{eq:3A_correction_IR_mixed}
\end{align}
in the external momentum of the external $U(1)$ leg. Note, that in contrast to the situation where all three external legs are in the $U(1)$ subsector where we had a sum over all three momenta and linear IR divergences in each one, \eqnref{eq:3A_correction_IR_mixed} exhibits such an IR divergence \emph{only} for the external momentum of the external $U(1)$ leg (for which we have chosen $p_3$ above) but not for the other two. It must be noted, that this IR behaviour is present in every single graph of \figref{fig:1loop_3A_all}, not just their sum. 
This fact was actually not clear from the previous work Ref.~\cite{Armoni:2000xr}.
Moreover the comment in that paper that the IR divergent $U(1)$-$SU(N)$-$SU(N)$ result was ``exactly the same as in the $U(1)$-$U(1)$-$U(1)$ case'' was in fact quite misleading, as there clearly is subtle difference.

Concerning the (would-be) logarithmic IR divergence, the situation is similar: the only log that survives is the one corresponding to the $U(1)$ leg. But due to that leg, there is only a sine (and no cosine), and hence that term is in fact finite as well.

\subsubsection*{Checking consistency with the Slavnov-Taylor identities}
We start by checking \eqnref{eq:id-3} for the IR divergent terms appearing at one-loop order.
The lhs of this identity exhibits an IR divergence if either all three colour indices $A,B,C$ are 0, or only one of them.
In both cases, the second term on the rhs vanishes since $f^{D0C}=0$.
If all three are 0, the identity reduces to the exact same form as in the $U_\star(1)$-case.
If only $A=0$ and $B=b$, $C=c$, the lhs has a linear IR divergence for the $(A=0,\s,x)$-leg, while the rhs reduces to
\begin{align}
 \frac{\ig\sqrt{2}}{\sqrt{N}} \!\co{\vvar{\Gam}{A^c_\s(x)}{A^b_\n(y)}}{\d(y-z)}+\ig d^{Dbc}\!\co{\vvar{\Gam}{A^D_\n(y)}{A^0_\s(x)}}{\d(x-z)}
 \,,
\end{align}
and the IR divergence comes only from the $D=0$ contribution reading
\begin{align}
 \frac{\ig\sqrt{2}}{\sqrt{N}} \d^{bc}\!\co{\vvar{\Gam}{A^0_\n(y)}{A^0_\s(x)}}{\d(x-z)}
 \,,
\end{align}
which is consistent.
If, on the other hand, only $C=0$ and $A=a$, $B=b$, the lhs is IR finite because the IR divergent term in the $(C=0,\m,z)$-leg is killed by the $z$-derivative due to transversality.
The rhs in that case reduces to
\begin{align}
 \frac{\ig\sqrt{2}}{\sqrt{N}} \!\co{\vvar{\Gam}{A^a_\n(y)}{A^b_\s(x)}}{\d(x-z)}
 \,,
\end{align}
which is IR finite as well (so consistent once more).

\subsection{\texorpdfstring{$4A$}{4A}-vertex corrections}

Another question to ask, is whether IR divergences are absent when all external legs are $SU(N)$ also when considering one-loop corrections to the $4A$-vertex. This has not previously been clear. For example, \cite{Armoni:2000xr} merely states that ``the calculation of these vertices is straightforward, though tedious''. 
\begin{figure}[!ht]
 \centering
 \includegraphics[scale=0.8]{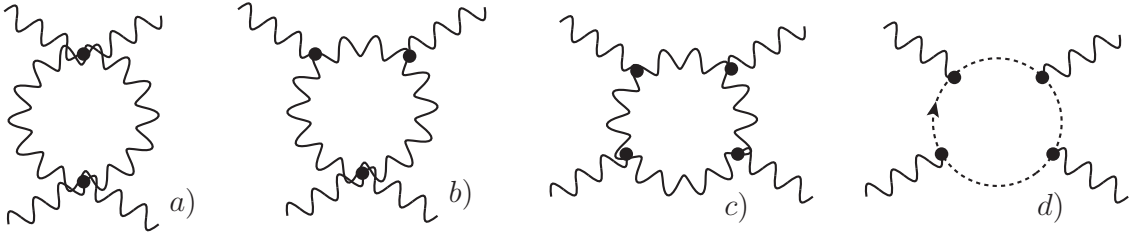}
 \caption{One loop corrections to the 4A-vertex.} 
 \label{fig:1loop_4A_all}
\end{figure}

In fact, explicit calculations show that two types of logarithmic IR divergences may be present in a graph, as we shall now explore by means of the simplest example of the graph with internal ghost loop (cf. \figref{fig:1loop_4A_all}d) with $N=2$ and $\a=1$:
if one considers all four external legs to be in the $U(1)$, one finds
\begin{align}
\Gamma^{4A^{u1}\!,\text{gh}}_{\textrm{IR},\m\n\r\t}\big(p_{1-4}\big)&=-\frac{\pi ^2 g^4 }{48}\left(\d_{\m\t} \d_{\n\r}+\d_{\m\r} \d_{\n\t}+\d_{\m\n} \d_{\r\t}\right) \nn\\
 &\quad\times\!  \bigg(\!\cos\! \left(\tfrac{\sth}{2} \left(p_1 \p_3+p_2\p_4\right)\right)\!
   \left(\ln\left(\p_2+\p_3\right)^2+ \ln\left(\p_1+\p_3\right)^2-\ln\p_1^2-\ln\p_2^2\right)
\nn\\
&\quad\quad +\cos \left(\tfrac{\sth}{2} \left(p_1 \p_2+p_3 \p_4\right)\right) \ln\left(\frac{(\p_1+\p_2)^2}{\sth^2\p_3^2\p_4^2}\right)\bigg)
\,. 
\end{align}
On the other hand, if all four external legs are in the $SU(2)$, the graph exhibits the infrared divergent contribution
\begin{align}
\Gamma^{4A^{su2}\!,\text{gh}}_{\textrm{IR},\m\n\r\t}\big(p_{1-4}\big)&=-\frac{\pi ^2 g^4 }{48} \left(\d_{\m\t} \d_{\n\r}+\d_{\m\r} \d_{\n\t}+\d_{\m\n} \d_{\r\t}\right) \nn\\
 &\quad\times\!  \Big(\!\cos\! \left(\tfrac{\sth}{2} \left(p_1 \p_3+p_2\p_4\right)\right)\!
   \left({\d }_{a d} {\d }_{b c} \ln\!\left(\sth^2(\p_2+\p_3)^2\right)+{\d }_{a c}
   {\d }_{b d} \ln\!\left(\sth^2(\p_1+\p_3)^2\right)\!\right) \
\nn\\
&\quad\quad+\cos \left(\tfrac{\sth}{2} \left(p_1 \p_2+p_3 \p_4\right)\right) {\d}_{ab} {\d}_{cd} \ln\left(\sth^2(\p_1+\p_2)^2\right)\Big)
\,, 
\end{align}
which is logarithmically divergent for vanishing of at least two external momenta. Notice, however, that only sums of two momenta appear in the logarithms, i.e. the $\ln \p_i^2$-terms which additionally appear for external $U(1)$ legs are absent here.
So there is indeed a difference between $4A$-vertex corrections with external legs in the $U(1)$ and those with external legs in the $SU(N)$, though the exact nature of this difference was not completely clear from \cite{Armoni:2000xr}.

However, the sum of all $4A$-vertex corrections must fulfill the Slavnov-Taylor identity \eqref{eq:id-4}.
Hence, all possible IR divergent terms are related to the ones of the $3A$-vertices, which by means of the Slavnov-Taylor identity \eqref{eq:id-3} are in turn related to the IR divergence \eqref{eq:vac-pol-IR-U1} and the log \eqref{eq:vac-pol-UV-U1}.
By power counting, the quadratic IR divergence of the vacuum polarization \eqref{eq:vac-pol-IR-U1} propagates to a linear IR divergence in the 3-vertex and is related to an IR-finite contribution in the 4-vertex.
Likewise, the log-terms of the vacuum polarization and 3-vertex propagate to the 4-vertex, leading to $\ln\p^2$-terms only for the external $U(1)$ legs.

Let us explore the nature of these terms some more:
It is  in fact known from the literature~\cite{Ruiz:2000}, that the IR divergence in the case where all legs are in the $U(1)$ is actually IR finite as can be seen from the appearing expressions
\begin{align}
\sin\left(\tfrac{\sth}{2} p_i\p_j\right)\sin\left(\tfrac{\sth}{2} p_k\p_l\right)\ln(\sth^2\p_i^2) \,,\nn\\
\sin\left(\tfrac{\sth}{2} p_i\p_j\right)\sin\left(\tfrac{\sth}{2} p_k\p_l\right)\ln(\sth^2(\p_i+\p_j)^2)
\,. 
\end{align}
For the same reason, the IR terms of the $U(1)$-$U(1)$-$SU(N)$-$SU(N)$ case are finite: The cosines drop out completely in that case, as can be deduced from \eqnref{eq:phase-factors}. 

When only one leg is $U(1)$, and the other are $SU(N)$, one has IR terms of the form
\begin{align}
f^{abc}\sin\left(\tfrac{\sth}{2} p_i\p_j\right)\sin\left(\tfrac{\sth}{2} p_k\p_l\right)\left(c_1\ln(\sth^2(\p_i+\p_j)^2)+c_2\ln(\sth^2\p_i^2)\right) \,,\nn\\
f^{abc}\sin\left(\tfrac{\sth}{2} p_i\p_j\right)\cos\left(\tfrac{\sth}{2} p_k\p_l\right)\left(c_1\ln(\sth^2(\p_i+\p_j)^2)+c_2\ln(\sth^2\p_i^2)\right)
\,,
\end{align}
which again are finite (i.e. no $\ln(\sth^2\p_k^2)$ terms are present). 

When all external legs are in the $SU(N)$, the only terms possible which might lead to IR-divergences would be of the form
\begin{align}
\cos\left(\tfrac{\sth}{2} p_i\p_j\right)\cos\left(\tfrac{\sth}{2} p_k\p_l\right)\ln(\sth^2(\p_i+\p_j)^2)
\,, 
\end{align}
but the Slavnov-Taylor identities \eqref{eq:id-3} and \eqref{eq:id-4} rule out the existence of any IR divergent terms in the 4-vertex with all external legs in the $SU(N)$.

\section{Comments on renormalizability}
\label{sec:comments}

In Ref.~\cite{Blaschke:2009e} it was argued, that renormalizability of a {\nc} gauge model could be restored by building an IR damping into the gauge propagator and implementing non-local counter terms for the quadratic and linear IR divergences using the ``soft-breaking'' techniques known from the Gribov-Zwanziger action~\cite{Gribov:1978,Zwanziger:1989,Zwanziger:1993,Baulieu:2009}. 
In a subsequent paper~\cite{Blaschke:2010ck} that idea was generalized to the non-Abelian case. 
However, it was overlooked, that for the mixed $U(1)$-$SU(N)$-$SU(N)$ vertex (cf. \eqnref{eq:3A_correction_IR_mixed}), only one counter term is needed (instead of a sum over all legs). 
This can easily be remedied by the replacement 
\begin{align}
\left(d^{ABC}-d^{abc}\d^{aA}\d^{bB}\d^{cC}\right)\aco{A^A_\m}{A^B_\n} \frac{\tilde{\pa}_\m\tilde{\pa}_\n\tilde{\pa}_\r}{\wsq^2}A^C_\r
\qquad \to \qquad 
\aco{A^A_\m}{A^A_\n} \frac{\tilde{\pa}_\m\tilde{\pa}_\n\tilde{\pa}_\r}{\wsq^2}A^0_\r
\end{align}
in the action (Eqn. (8) in that paper). 
Since $\var{A^0}{A^a}=0$ and $\var{A^A}{A^0}=\d^{0,A}$, this leads to the required counter terms for the 3A vertices. 

Furthermore, we have the Slavnov-Taylor identity \eqnref{eq:id-3} at our disposal, which we have shown to hold at least to one-loop order (for the IR divergences).
Hence, we will need one (Gribov-like) parameter less than was assumed in~\cite{Blaschke:2010ck}, namely $\g'\propto\g$.

The complete soft-breaking action of~\cite{Blaschke:2010ck} (i.e. the corrected Eqn. (8) of that paper) should hence read
{\allowdisplaybreaks
\begin{align}\label{eq:renormalizable_action}
\Act&=\Act_{\text{inv}}+\Act_{\text{gf}}+\Act_{\text{aux}}+\Act_{\text{soft}}+\Act_{\text{ext}}\,,\nonumber\\*
\Act_{\text{inv}}&=\intx\tinv{4}F^A_{\m\n}F^A_{\m\n}\,,\nonumber\\*
\Act_{\text{gf}}&=\intx\,s\left(\bc^A\,\pa_\m A^A_\m\right)=\intx\left(b^A\,\pa_\m A^A_\m-\bc^A\,\pa_\m (D_\m c)^A\right)\,,\nonumber\\
\Act_{\text{aux}}&=-\intx\,s\left(\bpsi^0_{\m\n}B^0_{\m\n}\right)=\intx\left(-\bB^0_{\m\n}B^0_{\m\n}+\bpsi^0_{\m\n}\psi^0_{\m\n}\right)\,,\nonumber\\
\Act_{\text{soft}}&=\intx\,s\Bigg[\!\left(\bar{Q}^0_{\m\n\a\b}B^0_{\m\n}+Q^0_{\m\n\a\b}\bB^0_{\m\n}\right)\inv{\wsq}\left(\!f^0_{\a\b}+\s\frac{\mth_{\a\b}}{2}\tilde{f}^0\!\right)
+Q'^0\aco{A^A_\m}{A^A_\n} \frac{\tilde{\pa}_\m\tilde{\pa}_\n\tilde{\pa}_\r}{\wsq^2}A^0_\r\Bigg]\nonumber\\*
&=\intx\bigg[\!\!\left(\bar{J}^0_{\m\n\a\b}B^0_{\m\n}+J^0_{\m\n\a\b}\bB^0_{\m\n}\right)\!\inv{\wsq}\!\left(\!f^0_{\a\b}+\s\frac{\mth_{\a\b}}{2}\tilde{f}^0\!\right)\! - \bar{Q}^0_{\m\n\a\b}\psi^0_{\m\n}\inv{\wsq}\!\left(\!f^0_{\a\b}+\s\frac{\mth_{\a\b}}{2}\tilde{f}^0\!\right)\nonumber\\*
& \qquad \qquad -\left(\bar{Q}^0_{\m\n\a\b}B^0_{\m\n}+Q^0_{\m\n\a\b}\bar{B}^0_{\m\n}\right)\inv{\wsq}\mathop{s}\left(\!f^0_{\a\b}+\s\frac{\mth_{\a\b}}{2}\tilde{f}^0\!\right)\nonumber\\*
&\qquad\qquad
+J'^0\aco{A^A_\m}{A^A_\n} \frac{\tilde{\pa}_\m\tilde{\pa}_\n\tilde{\pa}_\r}{\wsq^2}A^0_\r -
Q'^0s\left(\aco{A^A_\m}{A^A_\n} \frac{\tilde{\pa}_\m\tilde{\pa}_\n\tilde{\pa}_\r}{\wsq^2}A^0_\r\right)\!\bigg]\,, \nonumber\\
\Act_{\text{ext}}&=\intx\left(\W^A_\m (sA_\m)^A+\w^A (sc)^A\right)\,,
\end{align}
where 
}
\begin{align}\label{eq:def-eABC}
\wsq &= \tilde\partial_\mu \tilde\partial_\mu \,, &
f^0_{\m\n}&=\pa_\m A^0_\n-\pa_\n A^0_\m
\,, &
\tilde{f}^0&=\mth_{\m\n}f^0_{\m\n}
\,.
\end{align}
The multiplier field $b$ implements the Landau gauge fixing $\pa_\m A_\m=0$, $\bc$/$c$ denote the (anti)ghost, and $\s$ is a dimensionless parameter. The complex $U_\star(1)$ field $B^0_{\m\n}$, its complex conjugate $\bB^0_{\m\n}$ and the associated additional ghosts $\bpsi^0$, $\psi^0$ are introduced in order to implement the IR damping mechanism explained in Ref.~\cite{Blaschke:2009e} on the according $U_\star(1)$ gauge model. The additional $U_\star(1)$ sources $\bQ^0,Q^0,Q'^0,\bJ^0,J^0,J'^0$ are needed in order to ensure BRST invariance of the action in the ultraviolet. In the infrared they take the ``physical'' values
\begin{align}
&\bQ^0_{\m\n\a\b}\Big|_{\text{phys}}=Q^0_{\m\n\a\b}\Big|_{\text{phys}}=Q'^0\Big|_{\text{phys}}=0\,,  
&& J'^0\Big|_{\text{phys}}=-\frac{g\g^2}{2}\,,\nonumber\\
&\bJ^0_{\m\n\a\b}\Big|_{\text{phys}}=J^0_{\m\n\a\b}\Big|_{\text{phys}}=\frac{\g^2}{4}\left(\d_{\m\a}\d_{\n\b}-\d_{\m\b}\d_{\n\a}\right)\,,
\label{eq:physical-values}
\end{align}
where $\g$ is a Gribov-like parameter of mass dimension 1 (cf.~\cite{Gribov:1978,Zwanziger:1989,Zwanziger:1993,Baulieu:2009}). 
The action \eqref{eq:renormalizable_action} is hence invariant under the BRST transformations
\begin{align}\label{eq:BRST_of_renorm_action}
&sA_\mu=D_\mu c\,,  &&  sc=\ig {c}{c}\, ,\nonumber\\
&s\bc=b\,,          && sb=0\, ,  \nonumber\\
&s\bpsi_{\mu\nu}=\bB_{\m\n}\,,     && s\bB_{\m\n}=0\,,\nonumber\\
&sB_{\m\n}=\psi_{\m\n}\,,     && s\psi_{\m\n}=0\,,\nonumber\\
&s\bar{Q}_{\m\n\a\b}=\bar{J}_{\m\n\a\b}\,, && s\bar{J}_{\m\n\a\b}=0\,, \nonumber\\
&sQ_{\m\n\a\b}=J_{\m\n\a\b}\,, && sJ_{\m\n\a\b}=0\,,\nonumber\\
&sQ'=J'\,, && sJ'=0\,,
\end{align}
and for the non-linear transformations $sA_\m$ and $sc$, external sources $\W_\m$ and $\w$ have been introduced, respectively. 
For further details on this model, we refer to~\cite{Blaschke:2009e,Blaschke:2010ck}.
The above discussion indicates that the use of the full machinery of the Slavnov-Taylor identities will be determinant to explore the UV and IR sectors of the total action, as it was the case for instance in old studies on anomalies \cite{Ader:1987kh,Abud:1989up} or in the study of topological field theories for which the BRST symmetry must eventually be supplemented by conditions of equivariance-type to be implemented in the corresponding system of Slavnov-Taylor identities \cite{Wallet:1989,Baulieu:1990uv,Wallet:1995}.

\section{Two dimensional case}
\label{sec:2d}
The situation for the {\uim} is usually better in 2 dimensions.
In particular, the vacuum polarization tensor of the simplest {\nc} Yang-Mills action on $\mathbb{R}^2_\theta$ does not suffer from the hard as well as logarithmic IR singularities responsible for the mixing.
If one insists on using a ``covariant gauge'', then the theory is {\uim} free with the simple overall UV behaviour of a super-renormalizable field theory.
In fact, for the Yang-Mills theory on $\mathbb{R}^{2n}_\theta$, standard calculations show that the bad IR singular terms in the vacuum polarization tensor are proportional to $(D-2)$ (where $D=2n$) \cite{Blaschke:2008phd,wallet:jncg5}.
This cancellation propagates to other higher order correlation functions as a consequence of Slavnov-Taylor identities --- cf. \eqref{eq:id-3}, \eqref{eq:id-4}.
Alternatively, by choosing a ``temporal gauge'' as $A_2=0$, the gauge fixed action simply splits into a free gauge part and a free ghost part, akin to what happens in e.g. 2-dimensional QCD \cite{Frishman:1992mr}.

In 2 dimensions, it is known that massless field theories usually exhibit additional IR singularities.
In order to distinguish them from the other ones, we will sloppily call them ``2-dimensional IR singularities''.
In the case of 2-dimensional Yang-Mills theories on commutative spaces, these singularities obviously depend on the choice for the gauge fixing function.
While these singularities can of course be expected to have a similar dependence for the planar diagrams of the Yang-Mills theory on $\mathbb{R}^2_\theta$, it remains to examine the net IR behaviour when the corresponding non-planar diagrams are taken into account.
This is the purpose of the present section.
For the sake of clarity, it is convenient to begin with the case of a $U_\star(1)$ theory.

\subsection{\texorpdfstring{The $U_\star(1)$ case}{U(1)}}
In this subsection, we assume $\alpha=1$ for the gauge fixing parameter
and a rescaled $U(1)$ generator $T^0=1$.

\subsubsection*{Ghost 2-point function}
It is instructive to first exhibit the cancellation within the ghost 2-point function.
From the expressions for the ghost and gauge propagators and vertices, the 1-loop correction to the ghost 2-point function can be written as
\begin{equation}
\omega_g(p)=2g^2\int\! {\frac{d^Dk}{(2\pi)^D}}{\frac{1-\cos(\sth k\p)}{k^2(k+p)^2}}(p^2+pk)=2g^2p^2\int_0^1\!dx\,x\int\! {\frac{d^Dk}{(2\pi)^D}}{\frac{1-\cos(\sth k\p)}{(k^2+M^2)^2}}
\,, \label{omegaghost}
\end{equation}
($p$ is an external momentum) in which the planar and non-planar contributions have already been made apparent and $M^2:=p^2x(1-x)$.
By further using \eqnref{JN} of \appref{app:usefulintegrals}, we can write
{\allowdisplaybreaks
\begin{subequations}
\begin{align}
\omega_g(p)&=\omega_g^{P}(p)+\omega_g^{NP}(p)
 \,,\label{omegaghosttotal}
\\
\omega_g^{P}(p)&=2g^2p^2\int_0^1\!dx\,x{\frac{(M^2)^{{\frac{D}{2}}-2}}{ 
(2\pi)^{D/2}}}{\frac1{2^{D/2}}}\Gamma(2-{\frac{D}{2}})\nn\\*
&= {\frac{g^2 2^{1-{\frac{D}{2}} }}{(2\pi)^{D/2}(p^2)^{{\frac{D}{2}}-1} }} 
({\frac{\Gamma({\frac{D}{2}})\Gamma({\frac{D}{2}}-1) }{\Gamma(D-1) }})\Gamma(2-{\frac{D}{2}})
\,,\label{omegaghostplanaire2}
\\
\omega_g^{NP}(p)&=-2g^2p^2\int_0^1\!dx\,x{\frac1{ 
(2\pi)^{D/2}}}(M^2)^{{\frac{D}{2}}-2}\big( \frac12 ({\sqrt{ \sth^2\p^2M^2}})^{2-{\frac{D}{2}}}
K_{2-{\frac{D}{2}}}({\sqrt{ \sth^2\p^2M^2}})\big)
\,. \label{omegaghostnonplan}
\end{align}
\end{subequations}
For $D=2$, the UV finite planar contribution \eqref{omegaghostplanaire2} has an IR singularity stemming from the factor $\Gamma({\frac{D}{2}}-1)$ (see 2nd equality).
The UV finite non-planar contribution \eqref{omegaghostnonplan} also has an IR singularity coming from the factor $(M^2)^{{\frac{D}{2}}-2}=(M^2)^{-1}$.
The other potential source of IR singularity, $({\sqrt{ \sth^2\p^2M^2}})^{2-{\frac{D}{2}}}{{K}}_{2-{\frac{D}{2}}}({\sqrt{ \sth^2\p^2M^2}})$, which generates {\uim} for $D=4$ through the hard IR singularities $\sim 1/(\sth\p)^n$ stemming from \eqref{asympt1} is inoperative here since $z{{K}}_1(z)=1$ for $z\to 0$ (see 2nd relation in \eqref{relationbessel1}).
From  $K_1(z)\sim{\frac1{z}}+(az+bz^3+\ldots)+{\frac{z^2}{2}}\log(z)+\ldots$ ($a,\ b\in\mathbb{R}$), one obtains the small $|p|$ behaviour of \eqref{omegaghostnonplan}
}
\begin{equation}
\omega_g^{NP}(p)\sim-{\frac1{2\pi}}\int_0^1\!dx\,x{\frac1{x(1-x)}}+\ldots,
\qquad p\sim0
\,,\label{limnonplanghost}
\end{equation}
where the dots represent finite regular terms when $p\sim0$.
This IR singular term is exactly balanced by the planar contribution obtained from \eqref{omegaghostplanaire2} evaluated at $D=2$
\begin{equation}
\omega_g^{P}(p)={\frac1{2\pi}}\int_0^1\!dx\,x{\frac1{x(1-x)}}
\,.
\end{equation}
Hence \eqref{omegaghosttotal} is finite:
\begin{equation}
\lim_{p\to0}\omega_g(p)=\lim_{p\to0}(\omega_g^{P}(p)+\omega_g^{NP}(p))=\textrm{finite.}
\label{lighost2ptir}
\end{equation}

\subsubsection*{Vacuum polarization tensor}
The vacuum polarization tensor in $D$ dimensions may be written as
\begin{align}
 \Pi_{\m\n}&=\int\!d^Dk\, I_{\m\n}(k,p)\sin^2\!\left(\tfrac{\sth}{2}k\p\right)\nn\\
 &\approx \int\!d^Dk \left(I_{\m\n}(k,0)+p_\r\diff{}{p_\r}I_{\m\n}(k,p)\Big|_{p=0}+\frac{p_\r p_\s}{2}\ddiff{}{p_\r}{p_\s}I_{\m\n}(k,p)+\ldots\right)\sin^2\!\left(\tfrac{\sth}{2}k\p\right)
 \,,
\end{align}
where the first term exhibits the leading UV and related (through mixing) IR divergence.
It has been previously computed~\cite{Blaschke:2008phd,wallet:jncg5} that\footnote{In fact, it was shown that this leading divergence is \emph{independent of gauge fixing} ---  see also~\cite{Blaschke:2005b,Ruiz:2000}.}
\begin{align}
 I_{\m\n}(k,0)&\propto (D-2)\left(2\frac{k_\m k_\n}{k^2}-g_{\m\n}\right)
 \,, \label{eq:no-UV-in2d}
\end{align}
i.e. the leading hard divergences cancel in two dimensions.

The cancellation between the 2-dimensional IR singularities of the planar and non-planar parts of the vacuum polarization $\Pi_{\mu\nu}(p)$ again occurs as above, although the computation is a bit more involved.
$\Pi_{1\mu\nu}(p)$, $\Pi_{2\mu\nu}(p)$, $\Pi_{3\mu\nu}(p)$ correspond to the gauge loop diagram, the tadpole and the ghost loop diagram, respectively (see \figref{fig:vacpol_all}).
By using the expression for the vertices, a standard calculation yields
\begin{equation}
\Pi_{\mu\nu}(p)={\frac1{2}}\Pi_{1\mu\nu}(p)+{\frac1{2}}\Pi_{2\mu\nu}(p)-\Pi_{3\mu\nu}(p)
\,, \label{polartensor}
\end{equation}
with
{\allowdisplaybreaks
\begin{align}\label{omega1}
\Pi_{1\mu\nu}(p)&=2g^2\int\! \frac{d^Dk}{(2\pi)^D}\frac{1-\cos(\sth k\p)}{k^2(k+p)^2}
P_{\mu\nu}(p,k)
\,,\nn\\
P_{\mu\nu}(p,k)&=\delta_{\mu\nu}[(p-k)^2+(k+2p)^2]+k_\mu k_\nu(4D-6) \nn\\
&\quad +p_\mu p_\nu(D-6)+(2D-3)(p_\mu k_\nu+p_\nu k_\mu)
\,,\nn\\
\Pi_{2\mu\nu}(p)&=-4g^2\int\! \frac{d^Dk}{(2\pi)^D}\frac{1-\cos(\sth k\p)}{k^2(k+p)^2}(D-1)\delta_{\mu\nu}(k+p)^2
\,,\nn\\
\Pi_{3\mu\nu}(p)&=2g^2\int\! \frac{d^Dk}{(2\pi)^D}\frac{1-\cos(\sth k\p)}{k^2(k+p)^2}(k_\mu(k_\nu+p_\nu))
\,. 
\end{align}
When $D=2$, the UV finite planar contribution $\Pi^P_{\mu\nu}$ can be verified to be
}
\begin{equation}
\Pi^P_{\mu\nu}(p)={\frac{g^2}{\pi}}\int_0^1dx{\frac1{p^2x(1-x)}}(p^2\delta_{\mu\nu}-p_\mu p_\nu)\label{polarplan}
\end{equation}
where the 2-dimensional IR singularity is apparent. By using \eqnref{polartensor} and \eqref{omega1}, the UV finite non-planar contribution is expressed as
\begin{equation}
\Pi^{NP}_{\mu\nu}(p)=-g^2\int\! \frac{d^Dk}{(2\pi)^D}\frac{\cos(\sth k\p)}{k^2(k+p)^2}(2(2-D)[k^2\delta_{\mu\nu}-2k_\mu k_\nu]+4p^2\delta_{\mu\nu}-(D+2)p_\mu p_\nu)
\,. \label{nppolarisation}
\end{equation}
In view of \eqref{JMU}, the first two terms between brackets are IR singular, behaving as $\log|{\sth\p}|$ and $\frac{\p_\mu\p_\nu}{\p^2}$, respectively.
The latter leading IR singularity would be responsible for {\uim} but is canceled (together with the logarithmic singularity) by the overall factor $(2-D)$ as we recalled at the beginning of this section.
Setting $D=2$ and using \eqref{JN}, \eqref{quantities}, we hence obtain
\begin{equation}
\Pi^{NP}_{\mu\nu}(p)=-{\frac{g^2}{\pi}}\int_0^1dx{\frac1{p^2x(1-x)}}\big({\sqrt{M^2\sth^2\p^2}}{{K}}_1({\sqrt{M^2\sth^2\p^2}})\big)(p^2\delta_{\mu\nu}-p_\mu p_\nu)
\,. \label{polarnonplan}
\end{equation}
From \eqref{polarplan} and \eqref{polarnonplan}, it can be easily seen that the IR singular part of $\Pi_{\mu\nu}^{NP}$ exactly cancels the IR singularity in $\Pi_{\mu\nu}^P$.
Hence
\begin{equation}
\lim_{p\to0}\pi(p^2)=\textrm{finite,}\qquad \Pi_{\mu\nu}(p):=\pi(p^2)(p^2\delta_{\mu\nu}-p_\mu p_\nu)
\,. \label{ligauge2ptir}
\end{equation}
From \eqref{lighost2ptir} and \eqref{ligauge2ptir}, it appears that IR singularities of 2-dimensional origin in the 1-loop planar parts of ghost and gauge 2-point functions are annihilated by their non-planar counterparts.
The cancellation holds true for the 3 and 4-point functions as it can be seen by using the Slavnov-Taylor identities \eqref{eq:id-3}, \eqref{eq:id-4} (where in the $U(1)$ case the antisymmetric structure constants $f^{ABC}$ vanish, of course).

In general the 2-dimensional IR singularities are expected to depend on the gauge choice.
Consider for instance the 1-loop 2-point functions for a commutative 2-dimensional pure Yang-Mills theory computed successively in the Landau gauge and the temporal gauge.
Then, 2-d IR singularities show up in the 2-point functions in the Landau gauge while they are simply absent in the temporal gauge where interactions disappear.
The situation is a bit different in the present case since there are no remaining 2-dimensional IR singularities in the 2-point functions:
A cancellation still operates between planar and non-planar parts whenever the gauge choice leaves interactions.
This suggests that the absence of the 2-dimensional IR singularities seems likely not to depend on the gauge choice, as in some sense the theory would behave as if it was massive.
At a computational level, the cancellation can be understood by the decomposition
\begin{align}
 \sin^2\!\left(\tfrac{\sth}{2}k\p\right)&=\tinv2\left(1-\cos(\sth k\p)\right)
 \,,
\end{align}
appearing in the numerators of the amplitudes for the 2-point functions, in view of the minus sign between the planar and non-planar parts.

It is interesting to extend the above analysis by coupling a massless fermion, since a dynamical mass generation mechanism for the $A_\mu$ as in the Schwinger model can be expected.
The first possible gauge invariant coupling is built from $\nabla_\mu\psi=\partial_\mu\psi-\ig A_\mu\star\psi$ with
\begin{equation}
S^1_F=\int\! d^2x\, {\bar{\psi}}{\slashed{\nabla}}\psi\,,\qquad {\slashed{\nabla}}:=\sigma^\mu\nabla_\mu\,,\qquad \sigma^1=\begin{pmatrix} 0&1\\1&0\end{pmatrix},\qquad \sigma^2=\begin{pmatrix} 0&\ri\\-\ri&0\end{pmatrix}
\,,\label{fermionfunda}
\end{equation}
and matter gauge transformations $u\in U_\star(1)$, $\psi^u=u\star\psi,\ {\bar{\psi}}^u={\bar{\psi}}\star u^\dag,\ (\nabla\psi)^u=u\star\nabla\psi$. 
The vertex function{\footnote{As before, all the momenta are incoming and momentum conservation is understood.}} is
\begin{equation}
V_1(k_1,k_2,k_3)=-\ri\sigma^\mu\exp\left(\tfrac{\ri\sth}2k_3\k_2\right)
\,, \label{couplingfunda}
\end{equation}
where $k_3$ (resp. $k_2$) is the ${\bar{\psi}}$ (resp. $\psi$) momentum. 
The 2nd possible gauge invariant coupling is obtained from the covariant derivative $D_\mu\psi:=
\partial_\mu\psi-\ig[A_\mu,\psi]_\star$ with $(D_\mu\psi)^u=u\star D_\mu\psi\star u^\dag$, $\psi^u=u\star\psi\star u^\dag$ for any $u\in U_\star(1)$.
It yields
\begin{equation}
S^2_F=\int d^2x{\bar{\psi}}{\slashed{D}}\psi\,,\qquad  {\slashed{D}}=\sigma^\mu D_\mu\,, \qquad V_2(k_1,k_2,k_3)=-i\sigma^\mu\sin\!\big(\tfrac{\sth}{2}k_3\k_2\big)
\,, \label{fermadjoint}
\end{equation}
where the conventions for the momenta are the same as above.
Standard computation for each respective coupling yields the UV finite expressions
\begin{subequations}
\begin{align}
\Pi_{\mu\nu}^{1F}(p)&={\frac{g^2}{p^2\pi}}(p^2\delta_{\mu\nu}-p_\mu p_\nu)\label{polferm1},
\\
\Pi_{\mu\nu}^{2F}(p)&=\frac{g^2}{2}\int \frac{d^Dk}{(2\pi)^D}{\frac{1-\cos(\sth k\p)}{k^2(k+p)^2}}
\tr(\sigma_\mu({\slashed{k}}+{\slashed{p}})\sigma_\nu{\slashed{k}})
\,, \label{selfenerg}
\end{align}
\end{subequations}
where in \eqref{polferm1} only the planar diagram contributes since exponential factors in the vertices balance each other.
Besides, the non-planar part of \eqref{selfenerg} can be cast into the form
\begin{equation}
\Pi_{\mu\nu}^{2FNP}(p)=-{\frac{g^2}{4\pi}}\int_0^1\!dx\left((2-D)\delta_{\mu\nu}{\cal{M}}_0({\sqrt{ \sth^2\p^2M^2}})-2\sth^2\p_\mu\p_\nu{\cal{M}}_{-1}({\sqrt{ \sth^2\p^2M^2}})+\ldots) \right)
, \label{nonplanarferm2}
\end{equation}
where the dots denote regular terms and we used \eqref{JN}, \eqref{JMU}.
Setting $D=2$, we arrive at
\begin{equation}
\Pi_{\mu\nu}^{2FP}(p)=\frac{g^2}{2\pi}(\delta_{\mu\nu}-\frac{p_\mu p_\nu}{p^2})\,, \qquad
\Pi_{\mu\nu}^{2FNP}(p)={\frac{g^2}{2\pi}}\frac{\p_\mu\p_\nu}{\p^2}+\ldots
\,, 
\end{equation}
from which it is easy to verify that $\Pi_{\mu\nu}^{2F}=\Pi_{\mu\nu}^{2FP}(p)+\Pi_{\mu\nu}^{2FNP}(p)=\Pi_{\mu\nu}^{1F}(p)$ \eqref{polferm1}, because in 2-dimensions $\th_{\m\n}$ is proportional to the epsilon tensor and hence $\frac{\p_\mu\p_\nu}{\p^2}=(\d_{\m\n}-\frac{p_\m p_\n}{p^2})$.
In view of the expression for $\Pi_{\mu\nu}^{1F}(p)$, a pole is induced in the propagator for the gauge field $A_\mu$ $\sim(p^2+{\frac{g^2}{\pi}})^{-1}$ which receives a Schwinger mass $\mu={\frac{g}{\sqrt{\pi}}}$ which coincides with the value for the mass in the Schwinger model~\cite{Schwinger:1962tp} --- see also~\cite{Ardalan:2010qb,Armoni:2011pa} and references therein.

\subsection{\texorpdfstring{The $U_\star(N)$ case}{U(N)}}

If the external legs of the 2-point functions are $U(1)$, the situation is exactly the same as before (apart from a factor $N$).
Hence, we consider the case where the external legs are $SU(N)$.
In this case the phases of the 2-point functions reduce to $N\d^{ab}$ according to \eqnref{eq:SUN-phases-vacpol}, i.e. the graphs are purely planar and do not exhibit {\uim}.
Furthermore, due to \eqnref{eq:no-UV-in2d} no UV divergences appear, but in contrast to the $U(1)$ case, there does exist a new IR divergence related to the masslessness because there is no non-planar part to cancel it.
This 2-dimensional IR singularity has the same structure as in the planar part of the $U(1)$ case.

Concerning the vertex corrections, all graphs are UV finite by power counting and hence also free of {\uim} related IR divergences.
However, there will once more be 2-dimensional IR singularities present due to the masslessness of the model, as can be inferred from the Slavnov-Taylor identities \eqref{eq:id-3}, \eqref{eq:id-4}.

\section{Conclusion}

In this paper, we have clarified some properties of one-loop IR divergences in {\nc} non-Abelian gauge field theories which were not clear from previous literature, and have verified their consistency with the tree-level Slavnov-Taylor identity \eqref{eq:id-3}.
In addition we have made more explicit previous claims in the literature that only graphs with one or more external $U(1)$-legs lead to dangerous {\uim} terms at one-loop order.

Furthermore, if the Slavnov-Taylor identities hold to all orders for IR divergent terms, the number of independent Gribov-like parameters reduce to one, namely $\g$.
This point should be helpful when attempting a rigorous proof of renormalizability of the according soft-breaking model \eqref{eq:renormalizable_action}.

Finally, we have also discussed some properties which are special to the 2-dimensional case.

\subsection*{Acknowledgements}
Discussions with H. Steinacker, P. Vitale and M. Wohlgenannt at early stages of this work are gratefully acknowledged.
D.N. Blaschke is a recipient of an APART fellowship of the Austrian Academy of Sciences, and is also grateful for the hospitality of the theory division of LANL and its partial financial support.

\appendix
\section{Structure constants and identities}\label{app:identities}
Following~\cite{Armoni:2000xr} we consider
{\allowdisplaybreaks
\begin{align}\label{eq:generator-properties}
\Tr(T^AT^B)&=\inv{2}\d^{AB}\,, &
T^0&=\inv{\sqrt{2N}}\id_N\,, &
d^{AB0}&=\sqrt{\frac{2}{N}}\d^{AB}\,,
\end{align}
and
\begin{subequations}
\begin{align}\label{eq:identities-fd}
f^{ade}f^{bde}=N\d^{ab}\,, \qquad \qquad
d^{ade}d^{bde}&=\left(N-\frac{4}{N}\right)\d^{ab}\,,\\
f^{ade}f^{bef}f^{cfd}-f^{ade}d^{bef}d^{cfd}-d^{ade}f^{bef}d^{cfd}-d^{ade}d^{bef}f^{cfd}&=2N\left(1-\frac{3}{N^2}\right)f^{abc}\,,\\
d^{ade}d^{bef}d^{cfd}-d^{ade}f^{bef}f^{cfd}-f^{ade}d^{bef}f^{cfd}-f^{ade}f^{bef}d^{cfd}&=2N\left(1-\frac{3}{N^2}\right)d^{abc}\,,
\end{align}
(cf. Ref.~\cite{Armoni:2000xr} for a proof). 
From these identities it furthermore follows that
\begin{align}
d^{aDE}d^{bDE}&=N\d^{ab}\,,\\
f^{aDE}f^{bEF}f^{cFD}-f^{aDE}d^{bEF}d^{cFD}-d^{aDE}f^{bEF}d^{cFD}-d^{aDE}d^{bEF}f^{cFD}&=2Nf^{abc}\,,\\
d^{aDE}d^{bEF}d^{cFD}-d^{aDE}f^{bEF}f^{cFD}-f^{aDE}d^{bEF}f^{cFD}-f^{aDE}f^{bEF}d^{cFD}&=2Nd^{abc}\,.
\end{align}
\end{subequations}
In fact, one alternatively derives the identities
\begin{align}
f^{aDE}f^{bEF}f^{cFD}&=-f^{aDE}d^{bEF}d^{cFD}=-d^{aDE}f^{bEF}d^{cFD}=-d^{aDE}d^{bEF}f^{cFD}=\frac{N}{2}f^{abc} \,, \nn\\
d^{aDE}d^{bEF}d^{cFD}&=-d^{aDE}f^{bEF}f^{cFD}=-f^{aDE}d^{bEF}f^{cFD}=-f^{aDE}f^{bEF}d^{cFD}=\frac{N}{2}d^{abc}
\,, 
\end{align}
which will be more useful for explicit loop computations.

\section{Useful integrals}
\label{app:usefulintegrals}
The IR behaviour of the correlation functions can be conveniently extracted by making use of the following integrals given in e.g \cite{wallet:jncg5}:
\begin{align}
J_N({\p})&\equiv\int\! \frac{d^Dk}{(2\pi)^D}\frac{e^{\ri\sth k{\p}}}{(k^2+m^2)^N}=a_{N,D}{\cal{M}}_{N-{\frac{D}{2}}}(\sth m|{\p}|)
\,, \label{JN} \\
J_{N,\mu\nu}({\p})&\equiv\int\! \frac{d^Dk}{(2\pi)^D}{\frac{k_\mu k_\nu  e^{\ri \sth k\p}}{(k^2+m^2)^N}} \nn\\
&=a_{N,D}\big(\delta_{\mu\nu}{\cal{M}}_{N-1-{\frac{D}{2}}}(\sth m|{\p}|)-\sth^2\p_\mu\p_\nu{\cal{M}}_{N-2-{\frac{D}{2}}}(\sth m|{\p}|) \big)
\,, \label{JMU}
\end{align}
where
\begin{equation}
a_{N,D}={\frac{2^{-({\frac{D}{2}}+N-1)}}{\Gamma(N)\pi^{\frac{D}{2}}}}\,,\qquad {\cal{M}}_Q(m|{\p}|)={\frac1{(m^2)^Q}}(\sth m|{\p}|)^Q{{K}}_Q(\sth m|{\p}|)
\,, \label{quantities}
\end{equation}
in which ${{K}}_Q(z)$ is the modified Bessel function of second kind where $Q\in{\mathbb{Z}}$.
Recall its properties
\begin{equation}
{{K}}_{-Q}(z)={{K}}_Q(z)\,, \qquad  \lim_{z\to0}z^\nu{{K}}_\nu(z)=2^{\nu-1}\Gamma(\nu)\,,\quad \nu>0\label{relationbessel1}
\end{equation}
so that the following asymptotic expansion holds true:
\begin{equation}
{\cal{M}}_{-Q}(\sth m|{\p}|)\sim2^{Q-1}{\frac{\Gamma(Q)}{{(\sth^2\p^2)}^{Q}}}\,,\quad Q>0
\,. \label{asympt1}
\end{equation}
The usual strategy is to perform continuation of the expressions for the correlation functions to arbitrary $D$-dimension and then going back to $D=2$.


\end{document}